\documentclass[article,twocolumn,showpacs,superscriptaddress,prl]{revtex4}
\usepackage{epsfig}
\usepackage{amssymb}

\def\be{\begin{equation}}       \def\ee{\end{equation}}
\def\bea{\begin{eqnarray}}      \def\eea{\end{eqnarray}}

\begin{document}

\title{Realizing Three-Dimensional Artificial Spin Ice by  Stacking Planar  Nano-Arrays} 

\author{Gia-Wei Chern}

\author{ Charles Reichhardt}

\author{Cristiano Nisoli}

\affiliation{Center for Nonlinear Studies and Theoretical Division, Los Alamos National Laboratory, Los Alamos, NM 87545, USA}

\begin{abstract}
Artificial spin ice is a frustrated magnetic two-dimensional nano-material,  recently  employed to study variety of tailor-designed unusual collective behaviours. Recently proposed extensions to three dimensions are based on self-assembly techniques and allow little control over geometry and disorder. We present a viable design for the realization of a three-dimensional artificial spin ice with the same level of precision and control allowed by lithographic nano-fabrication of the popular two-dimensional case. Our geometry is based on layering already available two-dimensional artificial spin ice and  leads to an arrangement of ice-rule-frustrated units which is topologically equivalent to that of  the tetrahedra in a pyrochlore lattice. Consequently, we show, it exhibits a genuine ice phase and its excitations are, as in natural spin ice materials, magnetic monopoles interacting via Coulomb law. 
\end{abstract}
\maketitle


Spin ice materials, such as rare-earth pyrochlores and artificial spin ice, are magnetic systems in which  frustrated
interactions lead to complex (partial) orderings and unusual collective behaviours~\cite{bramwell01,castelnovo12,nisoli13}.
Magnetic ions in pyrochlore spin ice form a network of corner-sharing tetrahedra whose classical magnetic macro-spins minimize the local interaction energy by obeying the two-in-two-out ice rule proposed by Pauling for the proton orderings in water ice~\cite{ice}: hence the name, spin ice. It has recently been demonstrated that  elementary excitations over the disordered ice manifold  of spin ice materials are  emergent magnetic monopoles that fractionalize from local dipole excitations~\cite{castelnovo08,jaubert09}. 

The artificial counterparts of these natural materials, artificial spin ices~\cite{wang06,nisoli13}, are nanostructured two-dimensional (2D) arrays of single-domain ferromagnetic bars that behave like giant Ising spins. Collective behaviour of the nanomagnets can be controlled through appropriate choices of material, geometry and  array topology. Because of their nano-scale interaction energies ($\sim10^3-10^5$ K depending on the size of the nanomagnet and mutual spacing) they reveal, at  accessible temperatures, emergent features which  in natural materials are seen only at very low temperature. Following the pioneering work of Wang {\em et al.} on the two-dimensional square-ice array~\cite{wang06,nisoli07}, 
artificial spin ices have been proposed and studied in diverse types of physical systems~\cite{libal06,libal09,latimer13,trastoy13} and geometries such as honeycomb (kagome ice)~\cite{tanaka06,qi08,arnalds12,schumann10,moller09,chern11,nisoli10,zhang13}, 
brickwork~\cite{li10}, triangular~\cite{mol12,rodrigues13,zhang12}, and pentagonal lattices~\cite{chern13}.
A systematic approach for designing 2D arrays with emergent ice-type frustration has also been proposed~\cite{morrison13,chern13b}, and recently realized experimentally~\cite{gilbert13}.

Most of the experimental efforts in such artificial frustrated magnets  has understandably focused on  two-dimensional systems: even with  mature nanolithography, it remains a great challenge to integrate a full three-dimensional (3D) structure with oblique angles between nanobars such as the pyrochlore lattice. Recently an interesting realization of a 3D artificial spin ice on a opal-like lattice was realized via self-assembly techniques~\cite{mistonov13,zhukov06}, which unfortunately do not allow for control over the lattice geometry and connectivity.

Conversely, planar nano-structured arrays have offered great flexibility in nano-fabrication, which has been recently exploited to produce arrays of exotic geometries~\cite{gilbert13} or desired super paramagnetic behavior. 
Maintaining  this kind of manufacturing flexibility while producing an arrangement that captures the 3D spin ice behaviours would be ideal. One way to transport the convenience of 2D nano-lithography to 3D fabrication is by stacking 2D structures, thus building the 3D material layer by layer. Then, the essential issue becomes the theoretical design of such layered structure. Essentially that entails tuning   Ising interactions between nearest-neighboring (NN) nanobars to mimic spin ice frustration. Finally, it is necessary to verify  theoretically that such design would produce the desired ice manifold. 

In this manuscript we propose  a nano--fabrication oriented design for a multilayer artificial spin ice structure, topologically equivalent to the pyrochlore lattice. We demonstrate that the degrees of geometrical frustration can be controlled by gauging the interlayer spacing. In particular, there exists a critical spacing such that the nearest-neighbor spin-spin interactions is fully frustrated as in the case of pyrochlore spin ice. Through Monte-Carlo simulations we show that it  exhibits an ice manifold, as well as lower entropy antiferromagnetic phases.  We further show that the effective interaction between monopolar excitations in our 3D design follows a Coulomb law in the ice phase, as in natural dipolar spin ice on the pyrochlore lattice.

\begin{figure}
\includegraphics[width=0.9\columnwidth]{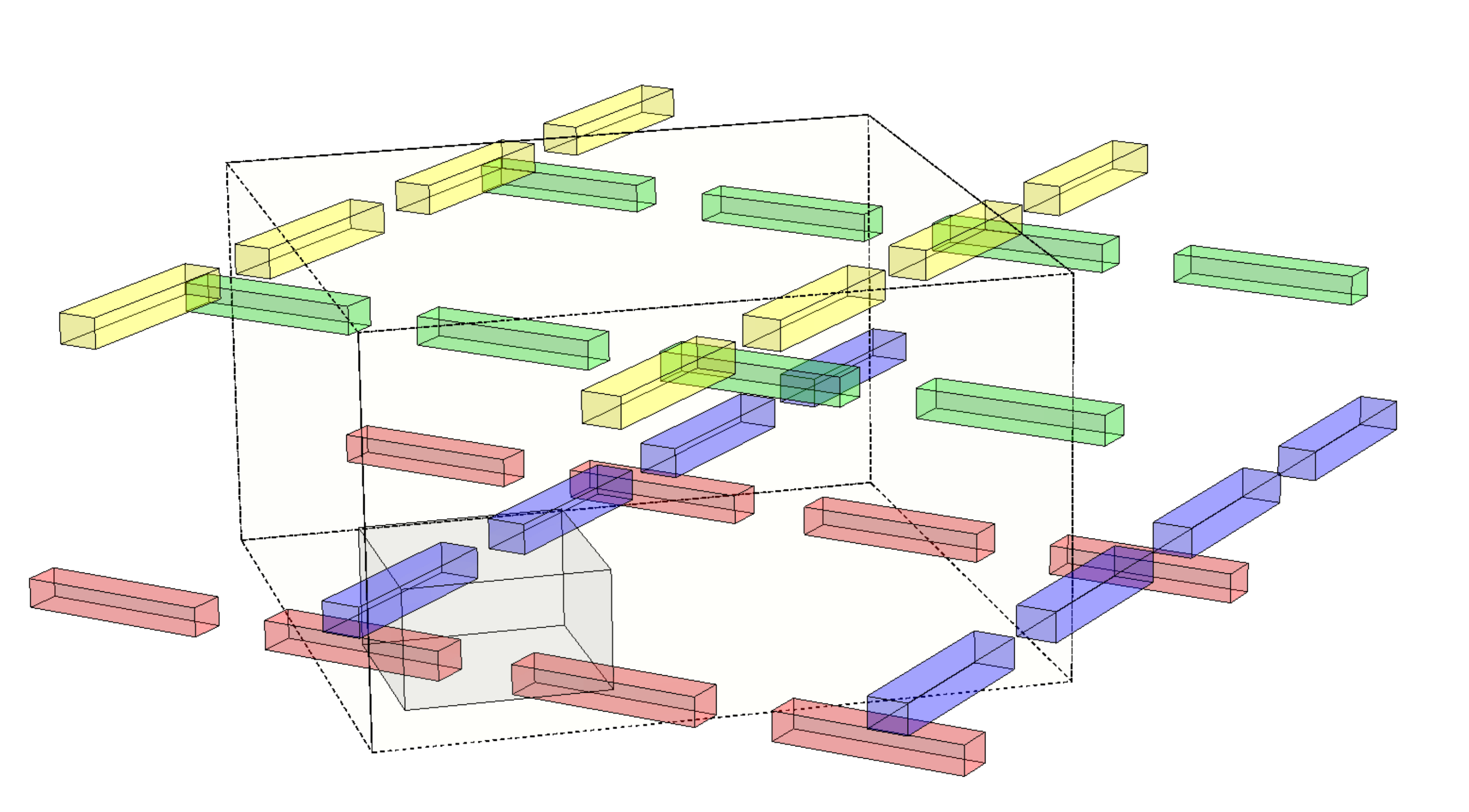}
\caption{\label{fig:layers} The multilayer structure of the proposed 3D artificial spin ice. Magnetic nano-islands
with different colors lie at different layers: red, blue, green,  and yellow bars lie at layers $z = 0$, $h$, $2h$, $3h$, respectively.  
The dashed big box indicates the conventional cubic unit cell in the corresponding pyrochlore
lattice. The shaded box with four nano islands at its corners indicates a frustrated unit, corresponding to a tetrahedron in the pyrochlore lattice. 
}
\end{figure}

Figure~\ref{fig:layers} shows the proposed multilayer structure. In each layer, parallel ferromagnetic bars
are arranged in a rectangular lattice with the long and short lattice constants being $2a$ and $a$, respectively;
the orientation of the nano islands is aligned with the short  axis. The arrays are rotated by 90$^\circ$
from one layer to the next. In addition, the arrays in every other layer are shifted by $a$ along the long axis.
The two-dimensional projection of the 3D structure on the $xy$ plane resembles the nano-magnetic arrays in a square ice.

\begin{figure}
\includegraphics[width=0.98\columnwidth]{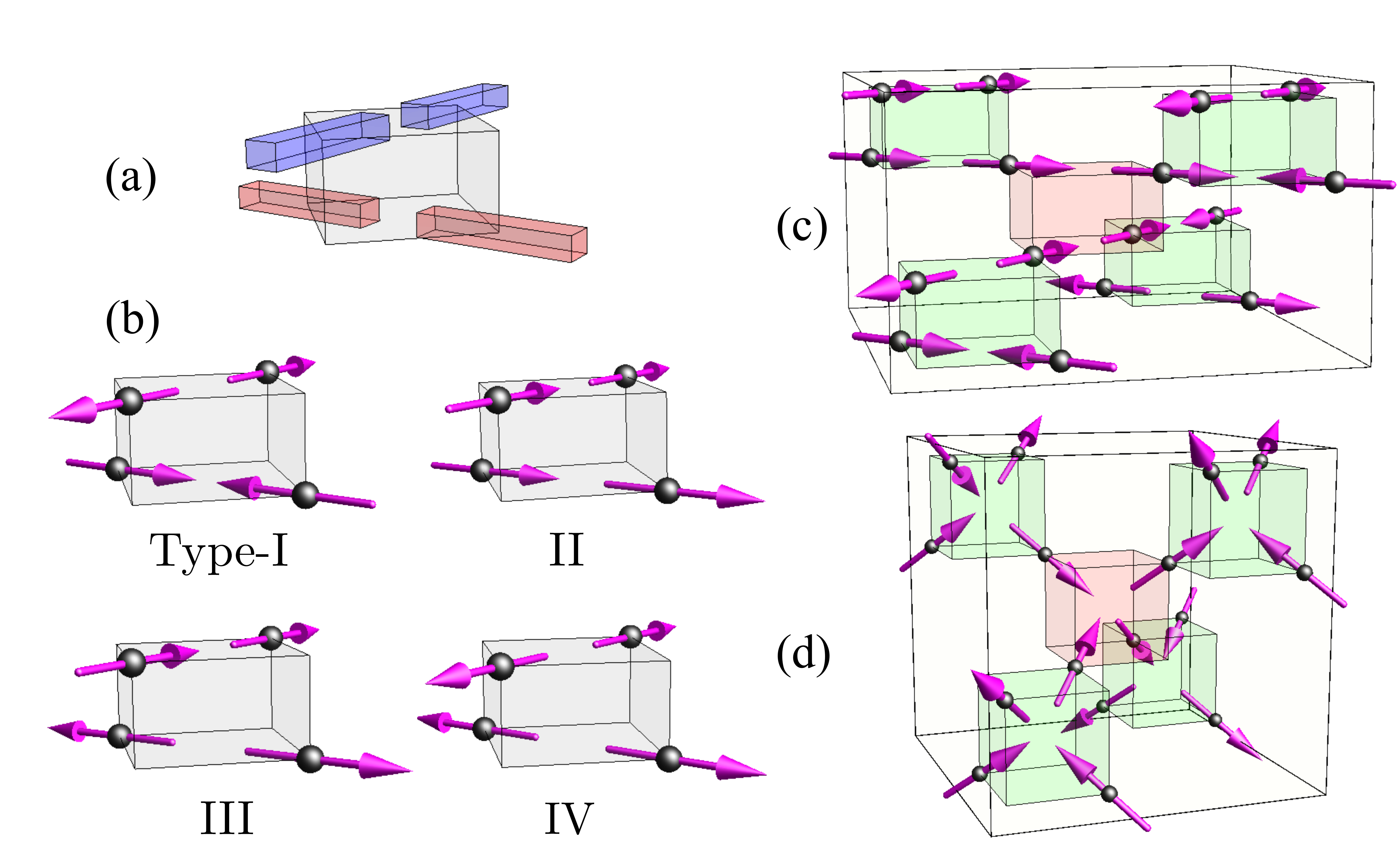}
\caption{\label{fig:ice3d} 
(a) The basic frustration unit in the multilayer artificial spin ice.
These units are analogous to vertices in square ice and tetrahedra in pyrochlore ice. 
(b) Ising-spin representation for the four distinct types of configurations in a tetrahedra.
(c) A random spin ice configuration for the magnetizations of nano-bars in the multi-layer structure. The corresponding
spin configuration in the pyrochlore lattice is shown in (d).}
\end{figure}

The basic frustration unit shown in Fig.~\ref{fig:ice3d}(a) consists of two pairs of nano-bars from consecutive layers.
The examples of Ising-spin representation of the magnetic state in the unit are shown in Fig.~\ref{fig:ice3d}(b). 
These units are analogous to the vertices and tetrahedra in 
square and pyrochlore ices, respectively.  Even though our design is a layered structure, the centers of these frustration units form a 3D lattice that is topologically equivalent to a diamond lattice. Moreover, as shown in Fig.~\ref{fig:layers}, each magnetic bar is shared by two frustration units, exactly as  each spin is shared by two tetrahedra in pyrochlore. These observations indicate that the nanomagnets themselves form a lattice that is isomorphic to pyrochlore, and each frustration unit corresponds to a tetrahedron. For convenience, we shall refer to the frustration unit in the multilayer spin ice as tetrahedron.
Fig.~\ref{fig:ice3d}(c) shows a configuration of random magnets satisfying the ice rules; the corresponding spin structure on the pyrochlore lattice is displayed in Fig.~\ref{fig:ice3d}(d).

Next we consider the energetics of the artificial spin ice. The isomorphism between our tetrahedra and those of a pyrochlore realization is clearly not enough: it is essential that the ice-rule is obeyed at the tetrahedron level. To this end, we first classify the tetrahedra into four types similar to vertices in a square ice~\cite{wang06}: type-I and II refer to 2-in-2-out units defined above, while type-III and IV denote the 3-in-1-out/1-in-3-out and all-in/all-out tetrahedra, respectively; see Fig.~\ref{fig:ice3d}(b).  
Assuming a single-domain magnetization for each nanobar~\cite{wang06}, the nanomagnets interact with each other through the dipolar interactions:
\begin{eqnarray}
	H = \frac{\mu_0}{8\pi}\sum_{i,j}
	\frac{\mathbf m_i \cdot \mathbf m_j - 3 (\mathbf m_i\cdot \hat{\mathbf r}_{ij}) (\mathbf m_j \cdot \hat{\mathbf r}_{ij})}{\left| \mathbf r_{ij} \right|^3},
\end{eqnarray}
where $\mathbf m_i = \pm \mu \hat{\mathbf e}_i$ is the dipole moment of $i$-th nanobar, and $\hat{\mathbf e}_i$ is a unit vector
parallel to the direction of the bar. 

The dipolar energy of the four nanomagnets in a unit reaches a minimum when they satisfy the ``ice rules", namely
two spins point toward the center and two point outward. 
However, degeneracy between the two types of 2-in-2-out tetrahedra, I and II in Fig.~\ref{fig:ice3d}(b), is in general lifted: unlike the equivalent pair-wise spin interactions in a tetrahedron,  the dipolar interaction energy between parallel bars (in the same layer) generally is different from that between orthogonal bars (in different layers), a situation 
similar to the case of square ice~\cite{wang06}. 

One can, however, restore the degeneracy between the types-I and II tetrahedra by properly adjusting the height $h$ of each layer, 
as pointed out in Ref.~\cite{moller06,mol10} for a two dimensional case.
For example, in the point-dipole approximation for the nanobars (with length $l \ll a$), equal dipolar energies for the six 2-in-2-out
configurations can be reached by setting $h = h^* = a \sqrt{(3/8)^{2/5} - 1/2} \approx 0.41890\,a$.
In practical realizations the interaction is not exactly dipolar and  the required height $h$ can be obtained with
the aid of micromagnetic simulations, accounting for the finite extension of the nanobars (see supplementary materials for two cases).
Away from $h^*$, the degeneracy is lifted and the lowest energy configurations are type-I (II) for $h < h^*$ ($h > h^*$).


We have thus established the equivalence of the multilayer artificial spin ice and the pyrochlore ice structurally and energetically, at least at the level of the nearest-neighbor interactions. We therefore expect our design to exhibit a similar ice manifold. 
Next we consider the thermodynamic transformations as a function of temperature for the 3D artificial spin ice.
Indeed, thanks to  recent advances in growth~\cite{morgan11} and thermal annealing technologies~\cite{zhang13, porro13}, it is now possible to systematically prepare artificial spin ice in thermal ensembles and even probe
its low entropy thermal states~\cite{porro13,morgan11,zhang13,nisoli12,greaves12,farhan13}.
Our numerical studies focus on multilayer structures with a height $h$ in the vicinity of $h^*$, estimated above, and obtain a phase diagram in the $h-T$ plane. For simplicity, we have used the point-dipole approximation for the magnetic nanobars. 
The long-range dipolar interactions are implemented using standard Ewald method,
and periodic boundary conditions were used in all simulations.

\begin{figure}
\includegraphics[width=1.0\columnwidth]{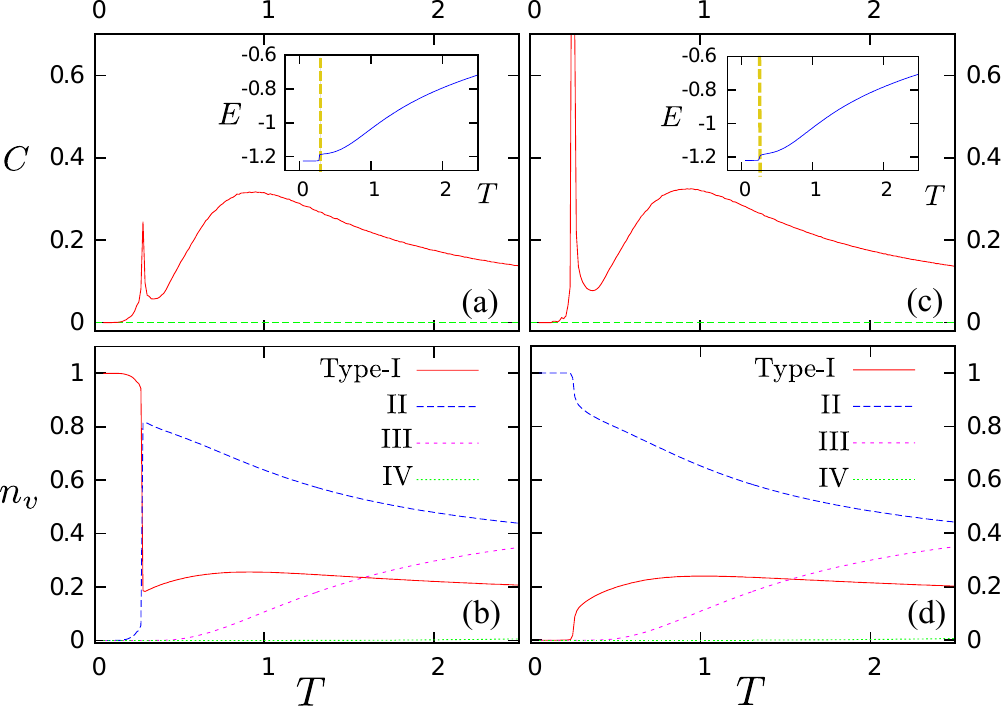}
\caption{\label{fig:sim1} Results of numerical Monte Carlo simulations with an interlayer spacing $h = h^*$ [panels (a),(b)], and $h = 1.013 h^*$ [panels (c),(d)]. 
The number of spins simulated is $N_s = 16\times 4^3$. 
(a) and (c) show specific heat $C$ as a function of temperature $T$, and the inset shows the energy density $E$ vs $T$. Here the temperature 
is measured in units of the nearest-neighbor dipolar interaction energy $D_{nn}$. (b) and (d) show the fraction of various tetrahedron types as a function of temperature.}
\end{figure}

The simulation results for the case with $h = h^*$ and $h = 1.013h^*$ are shown in Figs.~\ref{fig:sim1}. In both cases, the specific heat shows a broad peak at $T \approx D_{nn}$, where $D_{nn} = \mu_0\mu^2/2\pi a^3$ is the dipolar interaction energy between nearest neighbors, signaling the crossover into the two-in-two-out ice phase. 
Indeed, as shown in Fig.~\ref{fig:sim1}(b) and (d), population of monopoles (type-III tetrahedra) rapidly tends to zero below the crossover temperature (type-IV are always zero), where  most of the tetrahedra are in the 2-in-2-out states (type-I and II). 
As the temperature further decreases, the system undergoes a discontinuous transition at $T_N \approx 0.27 D_{nn}$ and $0.22 D_{nn}$, respectively for the two cases, revealed by sharp peaks in the specific heat. For the case with $h = h^*$, almost all tetrahedra in the ordered phase below $T_N$ are in the type-I state, similar to the case of square ice~\cite{farhan13,levis13}. On the other hand, the ordered state for $h = 1.013 h^*$ is predominantly composed of type-II vertices, indicating that a different ground state is selected for larger $h$.

\begin{figure}
\includegraphics[width=1.0\columnwidth]{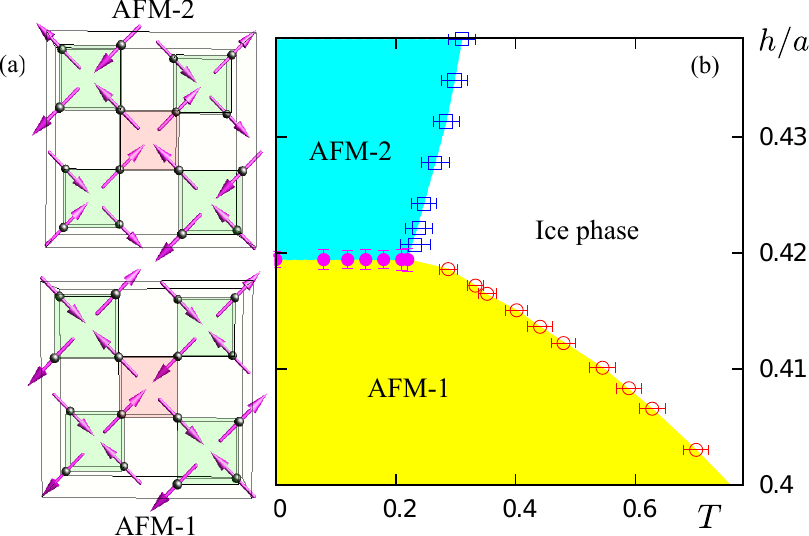}
\caption{\label{fig:phase} (a) Ground-state spin configurations for $h < h_c \approx 0.41943 a$ (AFM-1) and $h>h_c$ (AFM-2).
The phase diagram in the height-temperature plane is shown in (b).
}
\end{figure}

The apparent first-order magnetic transition at $T_N$ is induced by the dipolar interactions beyond the nearest-neighbor pairs,  a situation completely analogous to the low-temperature ordering of spins in the pyrochlore dipolar spin ice~\cite{melko01}. In fact, at $h = h^*$, the extensive degeneracy of the ice phase is already lifted
 by the second-nearest-neighbor interactions which lacks the pseudo-cubic symmetry of the NN pairs. When $h \neq h^*$, even the degeneracy at the NN level is lifted, as explained above. In order to obtain the 3D long-range spin order selected by the
dipolar interactions, we employed the loop algorithm~\cite{melko01} in combination with single-spin flips in our simulations to 
avoid the dynamical freezing of spins in the ice regime.  The phase diagram in the $h-T$ plane obtained
from our extensive Monte Carlo simulations is shown in Fig.~\ref{fig:phase}. We find two distinct long-range antiferromagnetic orderings at $T < T_N$. These two ordered phases are separated by a first-order line at $h_c \approx 0.4194 a$, in agreement  (only $10^{-4}$ larger) with the value $h^*$ at which the NN interactions are equivalent.

The ground state at $h \le h_c$ is characterized by a $\mathbf Q = \mathbf 0$ spin structure [AFM-1 in Fig.~\ref{fig:phase}(a)], implying a uniform ordering of type-I tetrahedra. This state is the 3D analog of the staggered arrangements of type-I vertices observed in the ground state of two-dimensional square ice~\cite{morgan11,zhang13}. In our 3D multilayer case, tetrahedra of different orientations are in the two different type-I configurations, respectively.

A different ground state was obtained when $h > h_c$. The second-neighbor interactions in this case
favor antiparallel alignment of magnetizations between nanobars of same orientation but  in different layers. 
The resultant 3D spin order, AFM-2 in Fig.~\ref{fig:phase}(a), consists of ferromagnetic (FM) layers of tetrahedra  stacked antiferromagnetically along the $z$ direction. As a result, each tetrahedron in the ground state has a type-II configuration with a net magnetization along the $x$ or $y$ direction, consistent with the temperature dependence of the tetrahedra fractions shown in Fig.~\ref{fig:sim1}(b). This magnetic structure is characterized by an ordering wavevector $\mathbf Q = (0, 0, 2\pi/c)$, where $c = 4h$ is the lattice constant in the $z$ direction. 
Interestingly, the same ordering was obtained in the dipolar pyrochlore spin ice~\cite{melko01}, 
where, however, the normal of the FM planes can point along either $x$, $y$, or $z$ directions due to the cubic symmetry. 

\begin{figure}
\includegraphics[width=.99\columnwidth]{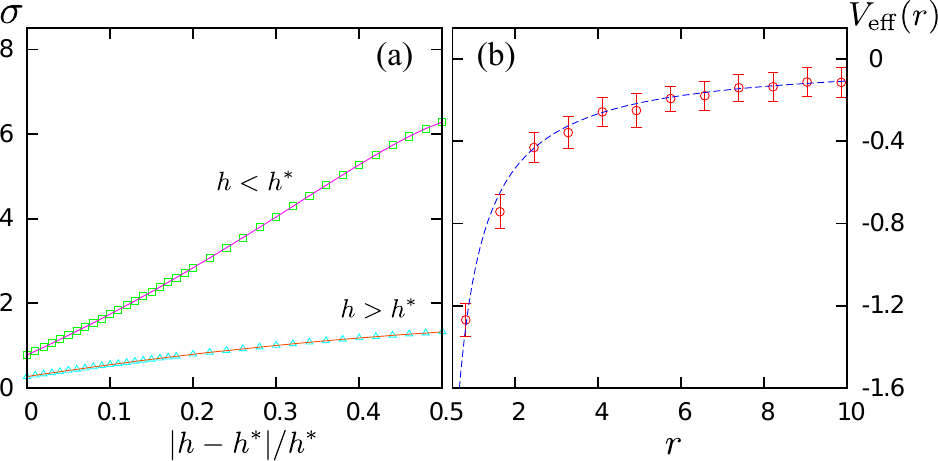}
\caption{
\label{fig:v_eff} (a) The tension $\sigma$ of a straight Dirac string connecting a pair
of monopoles in the two ordered states as a function of $h-h^*$. The two monopoles are separated by a displacement $\mathbf r = n (a/\sqrt{2}, 0, h)$ and $\mathbf r = n (0, a/\sqrt{2}, h)$, where $n$ is an integer, in the AFM-1 and 2 states, respectively. The distance $r$ is measured in units of $a$. 
(b) The effective potential $V_{\rm eff}(r)$ between a pair of monopoles (3-in-1-out and 1-in-3-out defects) separated by $\mathbf r = n (a/\sqrt{2}, 0, h)$.  The dashed line corresponds to $V_0/r$, with $V_0 \approx 1.3 D_{nn}$. 
}
\end{figure}

Next we study the low-energy elementary excitations in our 3D artificial spin ice. These are the type-III tetrahedra of Fig.~\ref{fig:ice3d}(b), and are endowed with a magnetic charge (magnetic monopoles~\cite{castelnovo08}). For $h \neq h^*$, creation of such a pair in the ordered phase introduces an array of excited 2-in-2-out tetrahedra along the Dirac string connecting the two monopoles. Indeed, as shown in Fig.~\ref{fig:v_eff}(a), the tension $\sigma$ of the Dirac string grows with increasing $|h - h^*|$ in the ordered states~\cite{mol10,mol09} (see supplementary materials for details). Interestingly, the tension is finite even when $h = h^*$ due to the long-range part of the dipolar interactions.


In the ice phase above $T_N$, the monopoles are deconfined as different ice configurations have approximately the same statistical weight. The effective interaction $V_{\rm eff}(r)$ between monopoles is thus expected to follow the Coulomb $1/r$ law. Assuming an ideal situation of exactly degenerate ice manifold, we numerically compute the effective potential by averaging over ice configurations that are compatible with two pinned monopoles at a distance $r$ in a system with $N_s = 16\times 8^3$ spins. As shown in Fig.~\ref{fig:v_eff}(b), a Coulomb law fits $V_{\mathrm{eff}}(r)$ well. This interaction is further screened due to the finite density of monopoles at finite temperatures~\cite{castelnovo08}. Consequently, the ice phase can be described by a plasma of weakly interacting magnetic monopoles. 

In summary, we have proposed a 3D layered geometry for a 3D artificial spin ice that captures the fully 3D spin ice behaviour such as an effective Coulomb interactions between monopoles and also provides an accessible and flexible, experimentally realizable geometry.
The structure is obtained by layering two dimensional lattices of nano-structures. The interest of this layered design lies in its accessibility via nano-fabrication of successive layers, and promises a viable way to extend artificial spin ice  to the third dimension.  

{\it Acknowledgment.}  We are grateful to J. Hollingsworth and S. Ivanov (Center for Integrated Nanotechnologies, LANL and Sandia) for useful discussions on  nano-fabrication techniques, and Cynthia Olson Reichhardt for a critical reading of the manuscript.
This work was carried out under the auspices of the National Nuclear Security Administration of the U.S. Department of Energy 
at Los Alamos National Laboratory under Contract No. DE-AC52-06NA25396.

\end{document}